\documentclass[11pt,twocolumn,superscriptaddress,longbibliography,reprint]{revtex4-1}

\usepackage{graphicx,amsmath,amssymb,color}
\usepackage{tabularx, ctable}
\usepackage{bm}
\usepackage{siunitx}
\usepackage{gensymb}

\usepackage[breaklinks=true,colorlinks,allcolors=blue]{hyperref}

\DeclareUnicodeCharacter{0301}{\'{e}}
\DeclareUnicodeCharacter{0301}{\'{0}}
\begin{document}

% \title{Non-Reciprocal Current-Phase Relation and Superconducting Diode Effect in \\ Planar Bi$_2$Se$_3$/NbSe$_2$  Josephson Junctions}

\title{Non-Reciprocal Current-Phase Relation and Superconducting Diode Effect in \\ Topological-Insulator-Based  Josephson Junctions}

\author{A. Kudriashov$^{+,*}$}
% \affiliation{Department of Materials Science and Engineering, National University of Singapore, 117575 Singapore}
\affiliation{Institute for Functional Intelligent
Materials, National University of Singapore, Singapore, 117575, Singapore}

\author{X. Zhou$^{+}$}
% \affiliation{Department of Materials Science and Engineering, National University of Singapore, 117575 Singapore}
\affiliation{Institute for Functional Intelligent
Materials, National University of Singapore, Singapore, 117575, Singapore}

\author{R.A. Hovhannisyan}
\affiliation{Department of Physics, Stockholm University, AlbaNova University Center, SE-10691 Stockholm, Sweden}

\author{A. Frolov}
\affiliation{Chemistry Department, M.V. Lomonosov Moscow State University, Moscow}

\author{L. Elesin}
\affiliation{Institute for Functional Intelligent
Materials, National University of Singapore, Singapore, 117575, Singapore}
\affiliation{Programmable Functional Materials Lab, Center for Neurophysics and Neuromorphic Technologies, Moscow 127495}

\author{Y. Wang}
\affiliation{Institute for Functional Intelligent
Materials, National University of Singapore, Singapore, 117575, Singapore}
\author{E.V. Zharkova}
\affiliation{Programmable Functional Materials Lab, Center for Neurophysics and Neuromorphic Technologies, Moscow 127495}

\author{T.~Taniguchi}
\affiliation{International Center for Materials Nanoarchitectonics, National Institute of Material Science, Tsukuba 305-0044, Japan}

\author{K.~Watanabe}
\affiliation{Research Center for Functional Materials, National Institute of Material Science, Tsukuba 305-0044, Japan}

\author{L.A. Yashina}
\affiliation{Chemistry Department, M.V. Lomonosov Moscow State University, Moscow}

\author{Z. Liu}
\affiliation{School of Materials Science and Engineering, Nanyang Technological University, Singapore, 639798, Singapore}

\author{Xin Zhou*}
\affiliation{Institute for Functional Intelligent
Materials, National University of Singapore, Singapore, 117575, Singapore}

\author{K.S. Novoselov}
\affiliation{Institute for Functional Intelligent Materials, National University of Singapore, Singapore, 117575, Singapore}

\author{D.A. Bandurin*}
\affiliation{Department of Materials Science and Engineering, National University of Singapore, 117575 Singapore}

% \author{authors}
% \affiliation{places}

\begin{abstract}
Josephson junctions (JJ) are essential for superconducting quantum technologies and searches of self-conjugate quasiparticles, pivotal for fault-tolerant quantum computing. Measuring the current-phase relation (CPR) in JJ based on topological insulators (TI) can provide critical insights into unconventional phenomena in these systems, such as the presence of Majorana bound states (MBS) and the nature of non-reciprocal transport. However, reconstructing CPR as a function of magnetic field in such JJs has remained experimentally challenging. Here, we introduce a platform for precise CPR measurements in planar JJs composed of NbSe$_2$ and few layer thick Bi$_2$Se$_3$ (TI) as a function of magnetic field. When a single flux quantum $\Phi_\mathrm{0}$ threads the junction, we observe anomalous peak-dip-shaped CPR behaviour and non-reciprocal supercurrent flow. We demonstrate that these anomalies stem from the edge-amplified sloped supercurrent profile rather than MBS signatures often invoked to explain puzzles emerging near $\Phi_\mathrm{0}$ in TI-based JJ. Furthermore, we show that such a supercurrent profile gives rise to a previously overlooked, robust and tunable Josephson diode effect.
These findings establish field-dependent CPR measurements as a critical tool for exploring topological superconducting devices and offer new design principles for non-reciprocal superconducting electronics.

\begin{center}

\end{center}
\end{abstract}

\maketitle
\section*{Introduction}

Characterizing and controlling supercurrent flow in Josephson junctions (JJs) is critical for advancing both fundamental research and practical applications, from superconducting classical and quantum technologies~\cite{Devoret,JJTHZ,SCComputer,KCSNSPDMW} to the discovery of exotic quasiparticles~\cite{Kitaev2001,Lutchyn,Feigelman2,Alicea2012,Beenakker,Kezilebieke2020,JoseLado}. 
While conventional methods, such as measuring the critical current between dissipationless and resistive states, have been instrumental in studying Josephson junctions (JJs), they often provide limited insight into fundamental mechanisms like spin-orbit coupling~\cite{SOCJJ}, quantum-geometric effects~\cite{QGEOMJJ}, and pairing symmetry~\cite{Pairing1} that govern their superconducting properties.
At the same time, spectroscopic techniques probing the amplitude of the superconducting wave function~\cite{Roditchev2015} lack direct access to phase-dependent phenomena such as current-induced hidden states~\cite{Yacoby} and screening currents in superconductor/ferromagnet hybrids~\cite{Robinson} that was uncovered only recently with the advent of sensitive non-invasive scanning probes. These limitations are amplified in JJs with topologically non-trivial weak links~\cite{williams2012,veldhorst2012,Ghatak2018,Feigelman1,williams2012,veldhorst2012,Ghatak2018}, where multiple confounding factors~\cite{Dynes1971,owen1967vortex,barone,hovhannisyan2024controllable,golod2022demonstration,golod2019two,foltyn2024quantum,rashidi2024,hovhannisyan2023superresolution} can mimic transport signatures of Majorana bound states~\cite{fu2008,potter2013,hegde2020} (MBS), pivotal for fault-tolerant quantum computing~\cite{Kitaev2001,Sau2010,nayak2008}.  Phase-sensitive information is also crucial for understanding non-reciprocal supercurrent transport and the Josephson diode effect (JDE), a research focus in superconducting nanoelectronics~\cite{nadeem2023}. While JDE has been observed in various systems~\cite{JJDE,Lyu2021,Ando2020,Reinhardt2024,nadeem2023,Robinson}, directly probing the direction and amplitude of non-reciprocal supercurrent in topological-insulator-based JJs as a function of external tuning knobs~\cite{Vic_KC}—such as a magnetic field—and uncovering its possible relation to MBS~\cite{MBS_diode,MBS-diode2} has remained experimentally difficult. These challenges have resulted in a proliferation of studies making unsubstantiated claims about MBS signatures  in some TIs and the nature of the JDE in related systems. To circumvent the limitations of the transport and spectroscopic approaches and get access to the internals of the topological JJs an alternative methodology framework is needed.

\begin{figure*}[ht!]
  \centering\includegraphics[width=\linewidth]{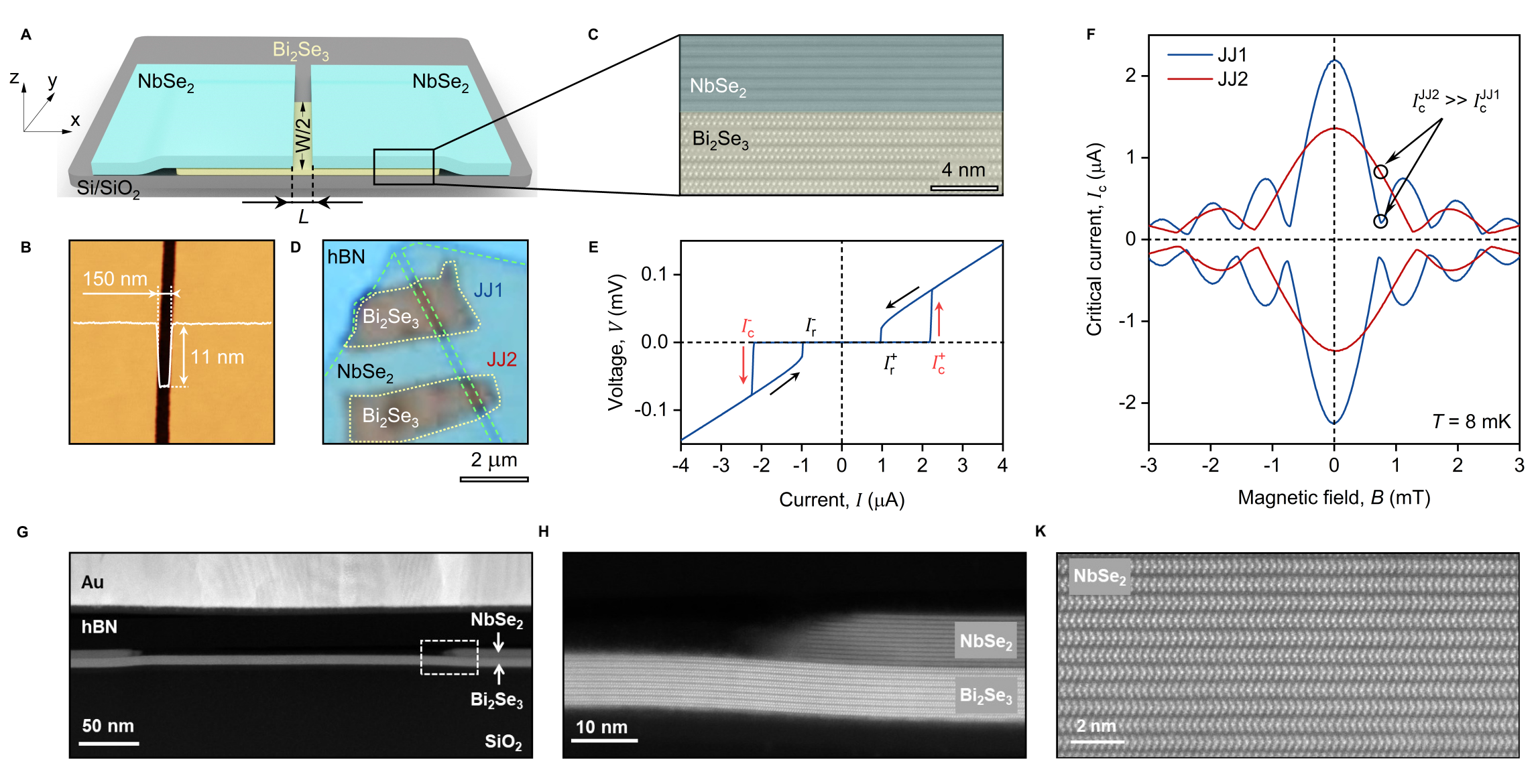}
\caption{\textbf{Planar VdW NbSe$_2$/Bi$_2$Se$_3$  Josephson junctions.}
\textbf{(A)} Schematic illustration of a single Josephson junction. A few-layer Bi$_2$Se$_3$ film is covered by a cracked NbSe$_2$ flake, with Ti/Au electrodes contacting the NbSe$_2$ regions. The device is fabricated on an oxidized Si substrate.
\textbf{(B)} AFM topography of the cracked NbSe$_2$ region.
White solid line indicates the height profile across the crack. Measurements performed in an Ar-filled glovebox.
\textbf{(C)} Representative high-resolution HAADF-STEM image across the NbSe$_2$/Bi$_2$Se$_3$ interface.
\textbf{(D)} Optical micrograph of an hBN-encapsulated NbSe$_2$/Bi$_2$Se$_3$ stack comprising two Bi$_2$Se$_3$ flakes that form two Josephson junctions (JJ1 and JJ2).
\textbf{(E)} Characteristic $I$-$V$ curve of JJ1 measured at base temperature $T=8$~mK. Arrows indicate the bias current sweep direction. $I^{+,-}_c$ and $I^{+,-}_r$ denote critical and retrapping currents for positive (+) and negative (-) current directions, respectively.
\textbf{(F)} Magnetic field dependence of the critical current $I_c$ for JJ1 (blue) and JJ2 (red) at the specified temperature.
\textbf{(G)} Typical STEM image of the device cross section. 
\textbf{(H)} Zoomed-in STEM image of the NbSe$_2$/Bi$_2$Se$_3$ interface at the end of the NbSe$_2$ flake. \textbf{(K)} STEM image of the atomic planes in NbSe$_2$.}
    \label{fig:F1}
\end{figure*}

In this work, we propose and realize such a framework that is based on accurate field-dependent reconstruction of the current-phase relation (CPR) in lateral JJ made of a superconducting NbSe$_2$ and one of the most-known vdW topological insulator (TI) - Bi$_2$Se$_3$~\cite{xia2009observation, qi2011topological}.
Surprisingly, although several studies have investigated the zero-field CPR of TI-based JJs~\cite{sochnikov2015,sochnikov2013,kayyalha2019,assouline2019,babich2023,endres2023}, none of them reported the exploration of the critical regime near $\Phi_0$, where MBSs are anticipated to dominate the supercurrent~\cite{potter2013,hegde2020}. At the same time, the CPR of conventional and topological JJs in the regime of JDE has remained largely unexplored. We fabricated lateral all-vdW heterojunctions and incorporated them into superconducting quantum interference devices (SQUIDs). By controlling the flux through the SQUID using a superconducting local flux line, we performed precise CPR measurements of individual JJs and its dependence on the external magnetic field. At $\Phi_0$, we observe unconventional CPR characteristics and non-reciprocal supercurrent flow near integer flux quanta, which we show to stem from the edge-amplified sloped supercurrent profile. The latter generates distinctive non-reciprocal CPR giving rise to a robust and tunable JDE with $30\%$ efficiency. Finally, we observe no signatures of MBS near a single flux quantum conditions in the CPR data, thereby challenging dozens of theoretical and experimental studies on supercondutor-proximitized Bi$_2$Se$_3$. 

\section*{Results} 

\textbf{Planar vdW Josephson junctions.} 
Our devices consist of planar Josephson junctions formed by two superconducting 2H-NbSe$_2$ electrodes coupled through a 5 nm thick topological insulator Bi$_2$Se$_3$ (Fig.~\ref{fig:F1}A)~\cite{mazumder2021brief}. 
Device fabrication was performed using a standard dry transfer method within an argon-filled glovebox to prevent surface degradation of the constituent materials~\cite{Cao2015}.
The fabrication process began with mechanical exfoliation of thin Bi$_2$Se$_3$ flakes onto a Si/SiO$_2$ substrate. 
Subsequently, an atomically-flat $\text{NbSe}_2$ flake, containing an intrinsic crack~\cite{zalic2023,dvir2021}, was transferred onto the Bi$_2$Se$_3$, forming two superconducting electrodes separated by a narrow gap of $L\approx150$ nm (Fig.~\ref{fig:F1}B). 
Unlike conventional thin-film deposition techniques that typically involve sputtering superconducting electrodes onto topological insulators—a process known to introduce structural and compositional disorder~\cite{kudriashov2022}—our method ensures atomically sharp interface between the materials critical for superconducting proximity~\cite{BiSNbSe2,yabuki2016,son2020}. 
High-resolution high-angle annular dark-field scanning transmission electron microscopy (HAADF-STEM) imaging (Figs.~\ref{fig:F1}C,G-K and Supplementary Mateirals) reveals the structural characteristics of the heterostructure, confirming the exceptional interface quality, which is important for Josephson junction experiments. 
The heterostructure was further covered by a relatively thick (40 nm) slab of hexagonal boron nitride (hBN) to protect the device during subsequent patterning using electron-beam lithography (See Methods and Supplementary Materials). 

\begin{figure*}[ht!]
  \centering\includegraphics[width=0.8\linewidth]{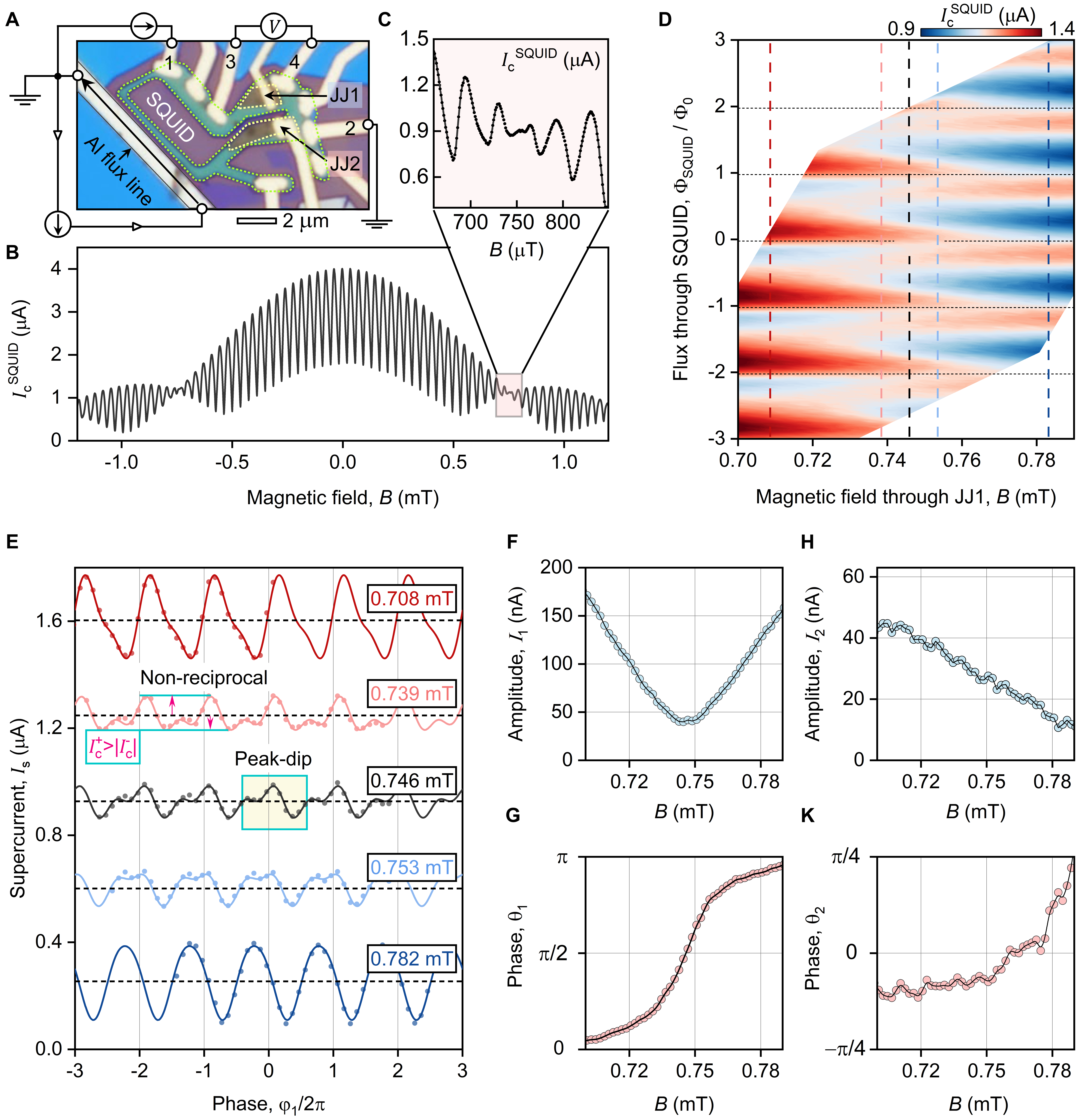}
 \caption{\textbf{NbSe$_2$/Bi$_2$Se$_3$ SQUID and CPR measurements.}
\textbf{(A)} Optical micrograph of the SQUID incorporating two Josephson junctions (JJ1 and JJ2) with measurement configuration overlay. DC current flows between contacts 1 and 2, while voltage is measured across contacts 3 and 4. A perpendicular magnetic field is applied externally, and a superconducting Al line controls the magnetic flux through the SQUID.
\textbf{(B)} Critical current $I^\mathrm{SQUID}_\mathrm{c}$ as a function of magnetic field, measured at base temperature $T=8$~mK.
\textbf{(C)} Magnified view of the region near one magnetic flux quantum through JJ1.
\textbf{(D)} $I^\mathrm{SQUID}_\mathrm{c}$ plotted against magnetic field ($B$) and magnetic flux through SQUID ($\Phi$), normalized to $\Phi_\mathrm{0}$.
\textbf{(E)} Current-phase relation (CPR) of JJ1 extracted along vertical dashed lines in panel (D).
Symbols represent experimental data; solid lines show best fits to $I^{JJ1}_s(\varphi_1)=I_0+I_1\sin{(\varphi_1+\theta_1)}+I_2\sin{(2\varphi_1+\theta_2)}$. Dashed lines represent $I_0$. Data is shifted vertically for clarity.
\textbf{(F)} Amplitude $I_1$ of the $1^\text{st}$ harmonic plotted versus magnetic field $B$.
\textbf{(G)} Phase $\theta_1$ of the $1^\text{st}$ harmonic plotted versus magnetic field $B$.
\textbf{(H)} Amplitude $I_2$ of the $2^\text{nd}$ harmonic plotted versus magnetic field $B$.
\textbf{(K)} Phase $\theta_2$ of the $2^\text{nd}$ harmonic plotted versus magnetic field $B$.
}
    \label{fig:F2}
\end{figure*}

We fabricated three different samples, all featuring robust proximity effect (Supplementary Information). 
The data in the main text is shown for one of them. This device contains two junctions, which we will refer to as JJ1 and JJ2, shown in the optical image in Fig.~\ref{fig:F1}D. 
We intentionally oriented NbSe$_\mathrm{2}$ crack in a way to ensure different junction widths and patterned the device in a SQUID geometry, which is required for CPR measurements (details below). 
While we first measured the as-fabricated SQUID, for convenience, we initially present data from the individual junctions, obtained after etching the SQUID into two independent devices. The \( I \)-\( V \) curve of the JJ1, shown in Fig.~\ref{fig:F1}E, is typical for superconductor-normal metal-superconductor Josephson junctions. 
It displays a zero-voltage state for currents below the critical current \( I_\mathrm{c} \), followed by a sharp transition to the resistive state as the current exceeds \( I_\mathrm{c} \).  
Upon decreasing the current, the junction remains in the resistive state until the current drops below the retrapping current \( I_r \), leading to hysteresis in the curve. 
Hysteresis is typical for this kind of device and is caused by overheating. 

Since the transition from superconducting to normal state is very sharp, it allows us to define a critical current $I_\mathrm{c}$ and a retrapping current $I_\mathrm{r}$ using the threshold method and accurately determine its dependence on the magnetic field,  $B$. 
Figure~\ref{fig:F1}F shows $I_\mathrm{c}$ as a function of $B$ for both JJ1 and JJ2.
It exhibits a characteristic Fraunhofer-like pattern, where the critical current \( I_\mathrm{c} \) oscillates as a function of the applied magnetic field \( B \), reaching minimum (non-zero) values when the magnetic flux through the junctions is close to integer multiples of the magnetic flux quantum, \( \Phi_{1,2} = n\Phi_0 \), where 1,2 indices correspond to JJ1 and JJ2 respectively and $n$ is integer.

\textbf{Current-phase relation measurements.} Since the widths of the two junctions were intentionally made different, their $I_\mathrm{c}(B)$  patterns also differ (as shown in Fig.\ref{fig:F1}~D,F). 
Importantly, when $B\approx0.75$ mT (which corresponds to the interference-pattern minimum), $I^\mathrm{JJ1}_\mathrm{c}\simeq60$~nA and $I^\mathrm{JJ2}_\mathrm{c}\simeq1~\mu$A. 
By redesigning the asymmetric-SQUID technique~\cite{CPR2006,CPR2007,babich2023} with an additional flux bias, we leveraged this order-of-magnitude asymmetry in critical current to enable measurements of the CPR of JJ1 as a function of external magnetic field, a capability previously unattained in TI-based JJs.

The idea behind this technique is the following. 
The supercurrent through the SQUID is the sum of supercurrents through individual JJs: $I^\mathrm{SQUID}_\mathrm{s}=I^\mathrm{JJ1}_\mathrm{s}(\varphi_\mathrm{1})+I^\mathrm{JJ2}_\mathrm{s}(\varphi_\mathrm{2})$, where $\varphi_\mathrm{1}$ and $\varphi_\mathrm{2}$ are the phase differences in the JJ1 and JJ2, respectively.
Assuming negligible inductance of the SQUID, the phase differences are related by the magnetic flux through the SQUID, $\Phi$, as $\varphi_\mathrm{1}-\varphi_\mathrm{2}=2 \pi \Phi/ \Phi_\mathrm{0}$. 
%and $\Phi_\mathrm{0}$ is the magnetic flux quantum. 
When the critical current of the JJ2 is significantly larger than that of the JJ1, $\varphi_2=\varphi^*$ where $\varphi^*$ is the phase at which the critical current of the JJ2 is achieved and it is almost independent of $\Phi$. As a result, the critical current of the SQUID can be expressed as
\begin{equation}
    I^\mathrm{SQUID}_c(\Phi)=I^{JJ2}_c+I^{JJ1}_s(\varphi_\mathrm{1}(\Phi))
\end{equation}
where $\varphi_1(\Phi)=\varphi^*+2\pi\Phi/\Phi_0$. Therefore, the desired CPR, $I^{JJ1}_s(\varphi_\mathrm{1})$, can be reconstructed from eq.~(1) through the accurate measurements of the $I^\mathrm{SQUID}_c(\Phi)$~\cite{babich2023}.

\begin{figure*}[ht!]
  \centering\includegraphics[width=0.85\linewidth]{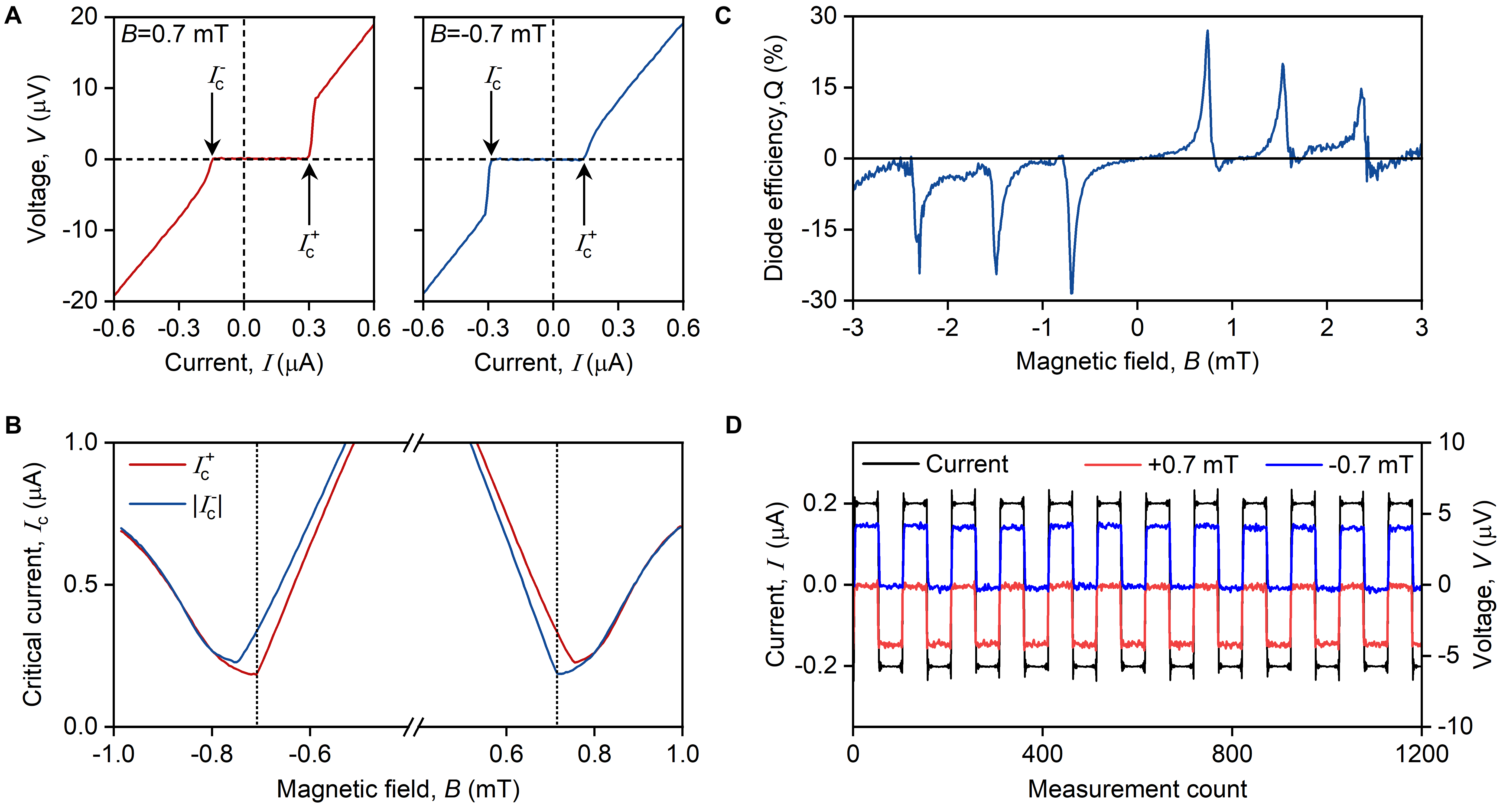}
    \caption{\textbf{Josephson diode in TI-based Josephson junctions.}
    \textbf{(A)} $I$-$V$ curves of the JJ1 measured at specified magnetic fields $B$. 
    \textbf{(B)} Positive ($I^+_c$) and negative ($I^-_c$) critical currents for JJ1 as a function of $B$ in the vicinity of one flux quantum through the junction. Vertical dashed lines correspond to $B=\pm0.7$~mT. 
    \textbf{(C)}  Superconducting diode efficiency, $Q$, as a function of $B$.
    \textbf{(D)} Rectified voltage across the JJ1 (right axis) measured by its excitation with the square-wave current (left axis) with the amplitude $0.2~\mu$A for $B=0.7$~mT (red curve) and $B=-0.7$~mT (blue curve). $T=8$~mK.
    } 
    \label{fig:F4}
\end{figure*}

To perform such experiment, we used a SQUID made of JJ1 and JJ2 shown in Fig.~\ref{fig:F2}A. Figure~\ref{fig:F2}B reveals fast $I^\mathrm{SQUID}_\mathrm{c}(B)$ oscillations modulated by the combination of interference patterns from individual JJs, highlighting the typical SQUID pattern. 
When $B\approx0.746$ mT, the amplitude of the oscillations is suppressed (Fig.~\ref{fig:F2}C), reflecting that JJ1 is close to the Fraunhofer minimum - the point of interest in our study. 
For an accurate control of $\Phi$ while maintaining magnetic field through the JJ1 constant, we endowed our device with an aluminum flux line located relatively far from the JJs (Fig.~\ref{fig:F2}A). By applying the current through the flux line and accounting for the superposition of both magnetic fields, we mapped the full dependence of $I^\mathrm{SQUID}_c$ on $\Phi$ and $B$ shown in Fig.~\ref{fig:F2}D.

Next, using this data and eq.~(1), we obtained the CPR $I^{JJ1}_s(\varphi_1)$ at selected values of $B$ in the vicinity of the first interference pattern minimum, as shown in Figs.~\ref{fig:F2}E. This is the central plot of our study. %The measured CPRs feature oscillatory pattern for all $\Phi_1$. 
At $B=0.708$~mT and $B=0.782$~mT, the CPRs are $2\pi$-periodic functions oscillating in anti-phase with respect to each other. 
This phase shift is expected since the JJs experience $0-\pi$ transition when crossing the integer number of flux quanta~\cite{hovhannisyan2024controllable}. 
At $B=0.746$~mT, the periodic pattern undergoes a notable transformation, developing an anomalous peak-dip structure (highlighted by the yellow rectangle) that deviates significantly from the anticipated CPR behavior where the critical current through JJ1 is expected to be fully suppressed (see below). 

% The observed peak-dip structure remarkably resembles the anomalous CPR due to MBS hybridizing at the edge of the junction predicted in Ref.~\cite{potter2013}. Moreover, the supercurrent in this regime is expected to be close to the maximal Josephson current carried by a single quantum channel $\Delta/\Phi_0\approx 75~$nA for the superconducting gap of $\Delta\sim1$~meV that is fortuitously close to our experimental value at the Fraunhofer minimum - 60 nA (see Fig.~\ref{fig:F2}e, black curve).

For further analysis of the observed anomalies, we characterize these features phenomenologically and fit the data with a two-harmonic expansion: 
$I^{JJ1}_\mathrm{s}(\varphi_1)=I_0+I_\mathrm{1}\sin{(\varphi_1+\theta_1)}+I_\mathrm{2}\sin{(2\varphi_1+\theta_2)}$ (solid lines in Fig.~\ref{fig:F2}E).
Here, $I_{1,2}$ and $\theta_{1,2}$ represent the amplitudes and phases of the first and second harmonics, respectively, and $I_0$ is the offset introduced by the asymmetric SQUID technique (see eq. (1)). 
The amplitude of the first harmonic $I_1$ decreases, reaching a minimum non-zero value at $B=0.746$~mT, before increasing again, while the phase $\theta_1$ undergoes a smooth $0-\pi$ transition, as shown in Figs.~\ref{fig:F2}F,G.
In contrast, the amplitude of the second harmonic $I_2$ gradually decreases over the entire measurement interval, showing no features associated with $B=0.746$~mT, while the phase $\theta_2$ remains close to zero, as shown in Fig.~\ref{fig:F2}H,K.
Therefore, there is a magnetic field dependent phase shift between first and second harmonics, which leads to another notable characteristic of the observed CPR: its directional asymmetry (the amplitudes of positive and negative supercurrents differ at certain magnetic fields, as marked in Fig.~\ref{fig:F2}E). 
This asymmetry implies that the junction exhibits a non-reciprocal transport, where the magnitude of the critical current depends on its direction through the JJ.

\textbf{Non-reciprocal CPR and Josephson diode.} 
To demonstrate and analyze the non-reciprocal behavior, we return to the data obtained on JJ1. Figure~\ref{fig:F4}A shows examples of two $I$-$V$ curves for the JJ1 measured at $B=\pm0.7$~mT and reveals the anticipated non-reciprocity: $I_\mathrm{c}^{+} \neq I_\mathrm{c}^{-}$ indicating the manifestation of the Josephson diode effect (JDE)~\cite{nadeem2023}.  Figure~\ref{fig:F4}B details this observation further by showing the $I_\mathrm{c}^{+}$ and $I_\mathrm{c}^{-}$ as a function of $B$ in the vicinity of the single flux quantum. 
The clear in-equivalence between the two persists for $|B|<0.8~$mT and abruptly disappears for larger $|B|$. 

For further analysis, we quantify the strength of the diode effect by introducing a diode efficiency factor 
$Q=(I^+_c-|I^-_c|)/(I^+_c+|I^-_c|)$. Figure~\ref{fig:F4}C shows the tunability of $Q$ on magnetic field B, revealing a distinctive tooth-like structure. The efficiency $|Q|$ exhibits periodic behavior, vanishing at integer numbers of flux quanta $\Phi_0$ through the junction, followed by sharp increases with increasing B. In our devices, the maximum achieved efficiency $|Q|$ reached approximately 30$\%$.

These large values of $|Q|$ allow us to demonstrate a robust rectification effect (Fig.~\ref{fig:F4}D). To this end, we applied square-wave current oscillations (black line) across the junction with the amplitude $0.2~\mu$A which is between $|I^-_c|=0.18~\mu$A and $I^+_c=0.3~\mu$A. The resulting voltage drop appears only during half of the excitation period, and its sign can be controlled by the direction of $B$ (cf. blue and red curves in Fig.~\ref{fig:F4}D).

\textbf{Theoretical modeling.} 
To understand the origin of the observed anomalous CPR and the JJ diode effect we use the standard expression~\cite{hovhannisyan2023superresolution,golod2022demonstration,goldobin2007josephson} that describes the supercurrent flow through the narrow JJ, i.e. when $W<\lambda_J$, where $W$ is the width of the junction and $\lambda_J$ is the Josephson penetration length (See Supplementary Information):
\begin{equation}\label{SingleJJ}
    I_s(\varphi_1) = \int_{-W/2}^{W/2} J_c(y) j_s\left(\alpha By + \varphi_1\right) \mathrm{d}y.
\end{equation}
Here $y$ is the coordinate along the junction with $y=0$ being the center of the junction, $J_c(y)$ is the supercurrent distribution (SD) at $B=0$, $j_s$ is the local current-phase relation, $\varphi_1$ is Josephson free phase, $\alpha = 2\pi L_\mathrm{eff}/\Phi_0$, and $L_\mathrm{eff}$ is the effective length of the junction, which is determined by the geometry of the superconducting leads in the case of a planar JJ~\cite{clem2010josephson}. 
Since $I_s(\varphi_1)$ describes the total supercurrent through the junction as a function of the phase difference between the superconducting leads $\varphi_1$, we will refer to this dependence as a global CPR, the desired property of the Bi$_2$Se$_3$/NbSe$_2$ JJ that we investigate in our study.
Equation~\ref{SingleJJ} allows to calculate the 
magnetic field dependence of the CPR as well as of the critical currents $I_\mathrm{c}^{+}=\max{I_\mathrm{s}(\varphi_1)}$ and  $I_\mathrm{c}^{-}=\min{I_\mathrm{s}(\varphi_1)}$.

\begin{figure*}[ht!]
  \centering\includegraphics[width=1\linewidth]{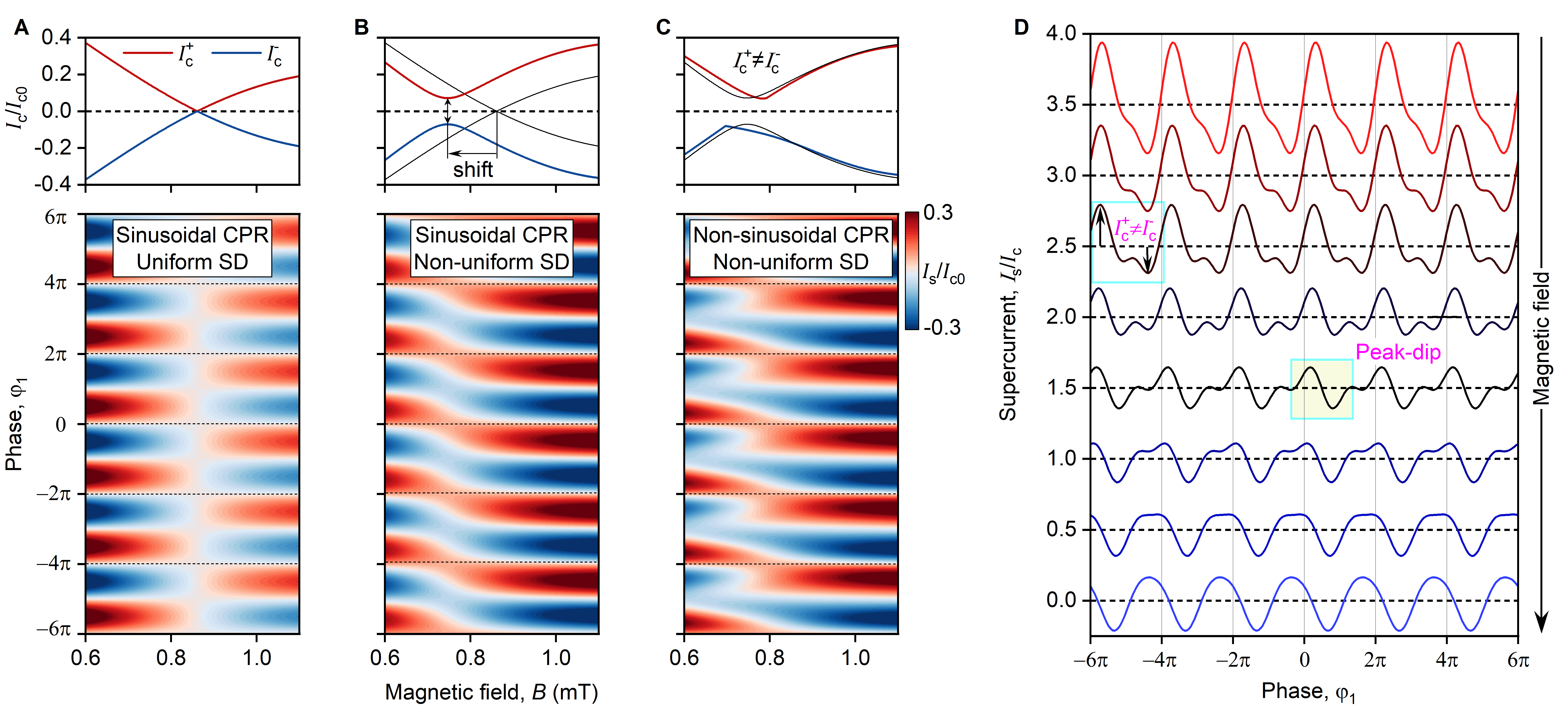}
    \caption{\textbf{Theoretical modeling of the CPR.}
    \textbf{(A-C)} Calculated positive $I_c^+$ and negative $I_c^-$ critical currents of the JJ as a function of magnetic field $B$ (upper panel) and calculated supercurrent $I_s$ as a function of phase difference $\varphi_1$ and magnetic field $B$ (lower panel). \textbf{A} The case of sinusoidal current-phase relation and uniform supercurrent distribution. 
     \textbf{(B)} The case of sinusoidal current-phase relation and non-uniform supercurrent distribution. Black line in the upper panel is the same as red and blue curves in (A).
     \textbf{(C)} The case of non-sinusoidal current-phase relation and non-uniform supercurrent distribution. Black line in the upper panel is the same as red and blue curves in (B).
    \textbf{(D)} Calculated $I_s(\varphi_1)$ for different magnetic fields $B$ around Fraunhofer minima for the case, shown in (C). The inset highlights the non-reciprocity of calculated CPR.
    } 
    \label{fig:F3}
\end{figure*}

First, we apply eq.~(\ref{SingleJJ}) to the simplest case of uniform supercurrent flow $J_\mathrm{c}(y)=1$ and sinusoidal local CPR $j_\mathrm{s}(\varphi)=\sin{(\varphi)}$. 
The results are shown in Fig.~\ref{fig:F3}A (lower panel) that maps $I_\mathrm{s}$ against $B$ and $\varphi_\mathrm{1}$.
The map reveals a sudden change of $I_\mathrm{s}$ sign at $\Phi_1=\Phi_\mathrm{0}$ that is achieved when $B=0.85~$mT demonstrating the standard $0-\pi$ transition. 
The calculated dependence is drastically different from the experimentally obtained CPR (cf. Fig~\ref{fig:F2}D) highlighting the distinctive supercurrent transport mechanism in our devices. 

The obtained $I_\mathrm{c}^{+,-}(B)$ dependencies, shown in Fig.~\ref{fig:F3}A (upper panel), describe conventional Fraunhofer pattern $I_c=I_{c0}|\text{sin}(\pi\Phi/\Phi_0)/(\pi\Phi/\Phi_0)|$ for ideal JJs which significantly deviate from experimentally measured $I_c(B)$ (Fig.~\ref{fig.ext}). 
This implies that our device features a non-uniform SD $J_\mathrm{c}(y)$ along the junction that needs to be accounted for in further analysis. 
To this end, we determine the SD $J_\mathrm{c}(y)$ using a standard method that calculates the Fourier transform of the measured $I_\mathrm{c}(B)$ dependencies~\cite{Dynes1971}. 
The results are shown by the red line in the inset of Fig.~\ref{fig.ext} that revels (i) increased $J_\mathrm{c}$ at the end of the JJ and (ii) the linear slope of $J_\mathrm{c}(y)$ across the whole junction. 

To take into account the non-uniform SD, for simplicity we approximate the extracted $J_\mathrm{c}(y)$ with a model dependence (black curve in Fig.~\ref{fig.ext}) and use it in eq.~(\ref{SingleJJ}).
The resulting $I_c(B)$ and $I_\mathrm{s}(B,\varphi_1)$ dependencies are shown in Fig.~\ref{fig:F3}B.
Compared with the previous case, non-uniform SD leads to the shift of the Fraunhofer minima, prevents the critical current from dropping to zero, and causes the global CPR to experience a gradual phase shift from 0 to $\pi$-state. However, accounting for non-uniform SD alone cannot explain the complex $I_\mathrm{s}(B,\varphi_1)$ pattern observed in Fig.~\ref{fig:F2}D. In particular, this model fails to account for the prominent second harmonic component present in the experimental global CPR (Fig.~\ref{fig:F2}E), suggesting an intrinsic non-sinusoidal $j_\mathbf{s}(\varphi_1)$ relation that we incorporate in our subsequent analysis.

Although the exact $j_\mathbf{s}(\varphi_1)$ relation for our JJ cannot be measured directly, since our SQUID is not in the asymmetric regime at zero $B$, we assume $j_s(\varphi)=I_1\text{sin}(\varphi)+I_2\text{sin}(2\varphi)$, where $I_1$ and $I_2$ are the amplitudes of the first and second harmonics, respectively. To illustrate how it affects the evolution of the global CPR, in Figs.~\ref{fig:F3}C,D we show the global CPR and $I_\mathrm{c}(B)$ results for $j_s(\varphi)=1.3\text{sin}(\varphi)-0.3\text{sin}(2\varphi)$. The resulting model accurately reproduces all experimentally observed features, namely: (i) non-reciprocal CPR (marked in Fig.~\ref{fig:F3}D), (ii) Josephson diode effect (Fig.~\ref{fig:F3}C, upper panel), (iii) gradual (non-sharp) $0-\pi$ transition (iv) peak-dip CPR structure close to the single flux quantum (yellow shaded area). The excellent agreement between our model and the experimental data confirms that the observed effects arise from the interplay between spatial distribution of the supercurrent, combined with non-sinusoidal $j_\mathbf{s}(\varphi_1)$ dependence. 

\section*{DISCUSSIONS}
\begin{figure}[ht!]
  \centering\includegraphics[width=0.9\linewidth]{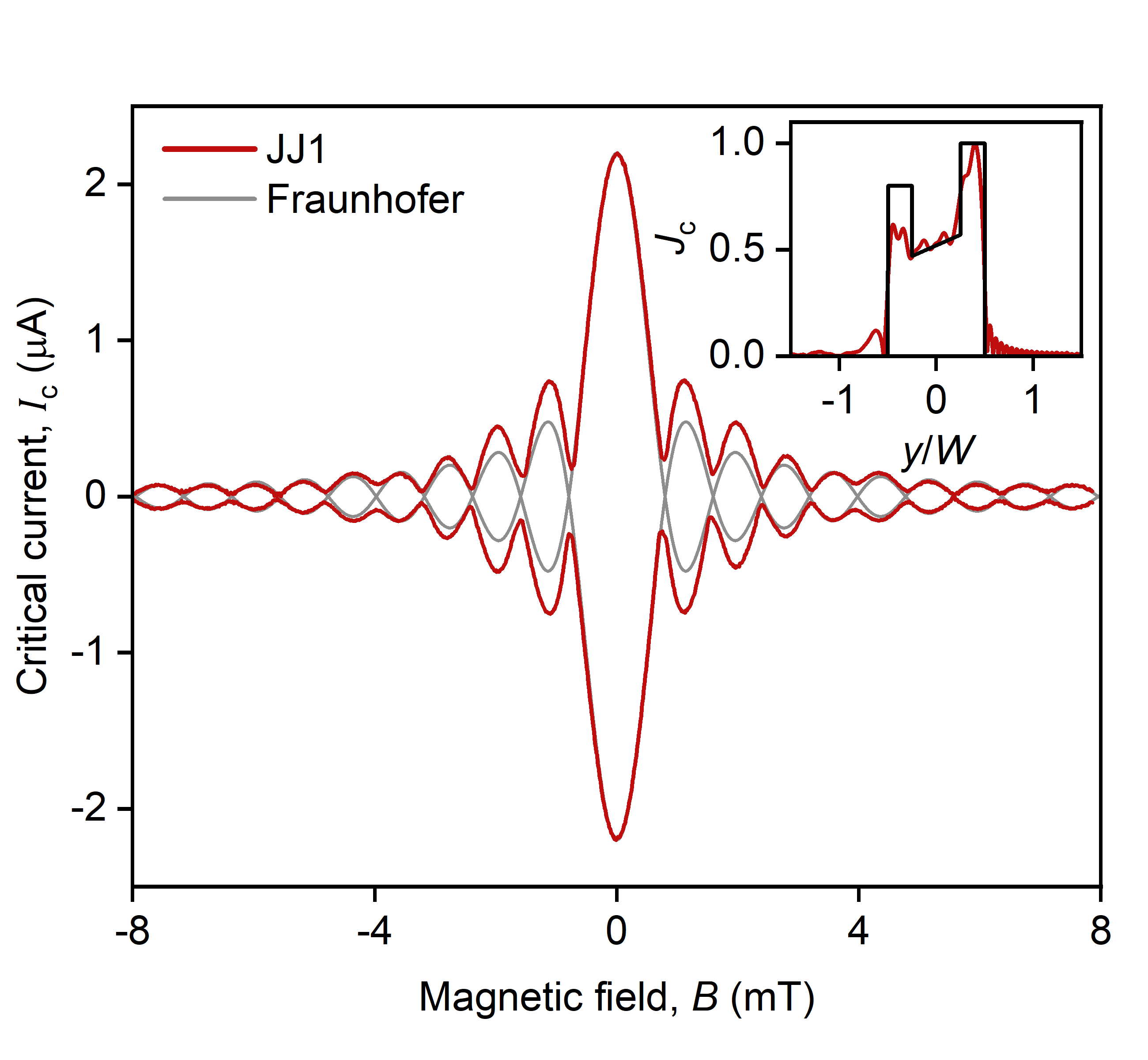}
    \caption{\textbf{Reconstruction of the supercurrent distribution.} 
    Critical current $I_c$ of JJ1 as a function of magnetic field $B$. The inset shows the supercurrent distribution, which was reconstructed from $I_c(B)$, by the red line, and the supercurrent distribution which was used for modeling, by the black line. 
    } 
    \label{fig.ext}
\end{figure}

Our theoretical analysis demonstrates that the observed anomalous CPR in our devices, its non-reciprocity and JJ diode effect can be explained by the interplay of non-uniform current distribution and higher-harmonic contributions.
A key prerequisite for the explanation is a distinct shape of $J_\mathrm{c}(y)$ shown in inset of Fig.~\ref{fig.ext}. The shape is characterized by (i) its linear gradient and (ii) increased $J_c$ at the ends of the JJ. Although the exact origin of the observed $J_\mathrm{c}(y)$ distribution remains unknown, we attribute these features to the following mechanisms. First, the slope can stem from a non-uniform gap between two superconducting $\text{NbSe}_2$ parts. Taking into account a steep dependence of the critical current on the distance between the superconductors, even a small asymmetry in the junction geometry can lead to the gradient in $J_c(y)$. Another possible mechanism involves Abrikosov vortices penetrating into the superconducting leads and creating non-uniform stray fields. However, trapped Abrikosov vortices would break time-reversal symmetry, violating the condition $I_c(B)=-I_c(-B)$ that is satisfied in our experiments excluding this interpretation. Second, the increased $J_c$ at the ends of the junction is likely related to current crowding. Indeed, in our devices, the width of the superconducting NbSe$_2$ leads exceeds the width of the $\text{Bi}_2\text{Se}_3$ flake, which can lead to the supercurrent streaming effect, thereby squeezing its density towards the edges. 

It is also instructive to place the observed JDE in a broader context of non-reciprocity in superconducting systems (See Ref.~\cite{nadeem2023} for recent review). Originally predicted to emerge in superconducting materials with intrinsically broken inversion and time-reversal symmetries, superconducting diodes can also be engineered artificially (e.g., Refs.~\cite{JJDE,Lyu2021,Ando2020,Yacoby,Reinhardt2024}). Among numerous approaches, recently proposed SQUID-based diodes particularly stand out in terms of ease of fabrication, large diode efficiency and remarkable tunability~\cite{souto2022josephson, fominov2022asymmetric}. The SD in our JJ1, namely its increase at the edges, effectively mimics that of a typical SQUID loop. Together with non-sinusoidal CPR and asymmetry between the edges, this gives rise to the JDE of a conceptually similar nature as in the proposed asymmetric higher-harmonic SQUIDs~\cite{souto2022josephson, fominov2022asymmetric}.

Finally, we compare our system with the model assumed by Potter and Fu in their original prediction of the peculiar CPR structure near the flux quantum due to MBS hybdridization~\cite{potter2013}. 
The theory considered a JJ based on a TI thin film in the short ($L<\xi_\mathrm{N}$) and narrow ($W<\lambda_J$) limit, where both top and bottom surfaces are proximitized by a superconductor, where $\xi_\mathrm{N}$ is the superconducting coherence length in Bi$_\mathrm{2}$Se$_\mathrm{3}$. 
Our junction satisfies these constraints, as detailed in the Supplementary Material. 
Furthermore, the thin $\text{Bi}_2\text{Se}_3$ flake ($\approx5$ nm) minimizes contributions from the side surfaces, while its high n-doping enables superconductivity to propagate from the top to the bottom surface through bulk states~\cite{BiSNbSe2}. 
Thus, our experimental system closely matches the conditions required by the theoretical proposal. The observed peak-dip structure (Fig.~\ref{fig:F2}E) in some sense resembles the anomalous CPR due to MBS hybridizing at the edge of the junction predicted in Ref.~\cite{potter2013} and can be naively attributed to their presence. Moreover, the supercurrent in this regime is expected to be close to the maximal Josephson current carried by a single quantum channel $\Delta/\Phi_0\approx 75~$nA for the superconducting gap of $\Delta\sim1$~meV that is fortuitously close to our experimental value at the Fraunhofer minimum - 60 nA (see Fig.~\ref{fig:F2}E, black curve). However, despite apparent similarity of the anomalous peak-dip CPR in the Potter-Fu model to the experimental data, there are two key differences that rule out this interpretation. First, the theoretical peak-dip structure is predicted to emerge around $\varphi_1=\pi$ but not at $\pi/2$ as in the case of our experiment. Second, the width of the experimental peak-dip structure is significantly larger compared to the theoretical prediction~\cite{potter2013}. 
Should the MBSs be present in the Bi$_2$Se$_3$-based JJs, the resolution of our technique would be sufficient to reveal their signatures.

% The width of the peak-dip structure was estimated as $2\pi\lambda_B/W$, where $\lambda_B$ is the Majorana decay length in y-direction~\cite{potter2013}. Phase shift. This phase shift can be explained by the fact that theory does not take into account the slope in the $J_c(y)$, which introduces additional phase shift, as demonstrated in Fig.~\ref{fig:F3}b.
% Second, the width of the peak-dip structure in the CPR should be $2\pi\lambda_B/W$, where $\lambda_B$ is the Majorana decay length in y-direction~\cite{potter2013}.
% %Based on our experimental data, this suggests that $\lambda_B/W=0.28$, which is significantly larger than the value considered in the original paper~\cite{potter2013}.  

To conclude, we have demonstrated a platform for precise CPR measurements in lateral Josephson junctions composed of NbSe$\textsubscript{2}$ and few-layer Bi$\textsubscript{2}$Se$\textsubscript{3}$, enabling the reconstruction of the CPR as a function of magnetic flux. 
Our findings reveal unconventional CPR characteristics and a robust, tunable Josephson diode effect in the vicinity of the flux quantum, stemming from edge-amplified supercurrent profiles and non-sinusoidal local current-phase relation. 
Notably, we find no evidence of Majorana bound states near a single flux quantum, challenging previous claims in Bi$\textsubscript{2}$Se$\textsubscript{3}$-based Josephson junctions. These results establish CPR measurements as a powerful tool for probing non-reciprocal transport phenomena in superconducting systems and provide new design principles for superconducting quantum devices.
It would be natural to extend our approach of CPR measurements in planar vdW Josephson junctions using cracked, atomically flat superconductors to investigate other topologically non-trivial materials such as Bi$_{1.1}$Sb$_{0.9}$Te$_2$S~\cite{Kushwaha2016}, WTe$_\mathrm{2}$~\cite{Sanfeng,WTe2JJ} or MnBi$_\mathrm{2}$Te$_\mathrm{4}$~\cite{MBT_QAH} that are unstable under standard nanofabrication processing.

\section*{Methods and Materials}

\subsection*{Sample fabrication}
Device fabrication was performed using a dry-transfer technique in an argon-filled glovebox with controlled H$_{2}$O and O$_{2}$ levels ($<0.5$ ppm). The fabrication process began with oxygen plasma cleaning of Si/SiO$_2$ substrates to optimize the subsequent exfoliation of thin Bi$_{2}$Se$_{3}$ flakes. Hexagonal boron nitride (hBN), NbSe$_{2}$, and Bi$_2$Se$_3$ were mechanically exfoliated onto the prepared Si/SiO${_2}$ substrates using adhesive tape. The heterostructure assembly proceeded by first picking up an hBN flake using a PDMS/PC (Polydimethylsiloxane/Polycarbonate) stamp, which was then used to pick up a cracked NbSe$_{2}$ flake, followed by transferring the assembled stack onto the exfoliated Bi$_2$Se$_3$ flake.

Electrical contacts were fabricated using standard electron beam lithography, followed by CHF$_3$/O$_{2}$ plasma etching of the hBN layer and electron beam evaporation of Ti/Au contacts (5 nm/65 nm). A magnetic flux line was subsequently created by depositing Ti (3 nm) followed by Al (97 nm). The SQUID loop geometry was then defined using CHF$_3$/O$_{2}$ reactive ion etching (RIE). After performing initial transport measurements on the complete SQUID, the device was divided into two separate Josephson junctions using CHF$_3$/O$_{2}$ RIE, enabling individual junction characterization after cooling.
See Supplementary information for details.

\subsection*{Scanning transmission electron microscopy}
A cross-sectional specimen of the nanodevice was prepared using a focused ion beam (FEI Versa 3D Dual Beam) after depositing a 30 nm-thick platinum layer on the surface as a protective coating. Cross-sectional STEM imaging was then performed ex situ using a JEOL ARM200F microscope equipped with an ASCOR aberration corrector, operating at 200 kV. Elemental mapping was conducted with an Oxford X-Max 100TLE EDS detector integrated with the microscope.

\subsection*{Low-noise measurements}

Electrical measurements were performed in a BlueFors dilution refrigerator at a base temperature of 8 mK, equipped with a 12~T superconducting solenoid. The solenoid was controlled using a Yokogawa GS610 source-measure unit. The devices were mounted on a QDevil QBoard sample holder using the BlueFors fast sample exchange system. The measurement setup incorporated two distinct wiring configurations to optimize different measurement requirements.

For precise $I$-$V$ measurements, we employed a low-current measurement line featuring a cryogenic filtering system consisting of a Basel Precision Instruments MFT25-100$\Omega$ microwave filter and thermalizer at the mixing chamber plate, complemented by a QDevil QFilter at the still plate. The setup was designed to effectively eliminate 50 Hz interference, allowing for fast and high-precision measurements. The devices were biased with a sinusoidal current (frequency: few Hz) generated by a QDevil qDAC-II and converted through a CS580 voltage-to-current converter. Both the sample voltage drop and applied voltage were amplified using a battery-powered SR560 amplifier and digitized synchronously with an NI-3232 system. Critical current determination was performed using custom LabView-based software (MeXpert) implementing a threshold method, with CS580 current offset compensation (See Supplementary information for details).

For high-current applications, specifically the magnetic flux line control requiring currents up to several mA, we implemented a separate configuration optimized to minimize heating effects. This setup utilized only a superconducting MFT25-25m$\Omega$ filter at the mixing chamber, bypassing the QFilter. The circuit employed aluminum superconducting bonds (additionally, QBoard resistors were removed), with current supplied by a QDevil qDAC-II through a 1k$\Omega$ resistor.

\setcounter{figure}{0}
\renewcommand{\thesection}{}
\renewcommand{\thesubsection}{E\arabic{subsection}}
\renewcommand{\theequation} {E\arabic{equation}}
\renewcommand{\thefigure} {E\arabic{figure}}
\renewcommand{\thetable} {E\arabic{table}}

% \begin{figure}[ht!]
%   \centering\includegraphics[width=\linewidth]{Extended TEM.JPG}
%     \caption{\textbf{Scanning transmission electron microscopy analysis of the vdW JJ. a,} Typical STEM image of the device cross section. \textbf{b, } Zoomed-in STEM image of the NbSe$_2$/Bi$_2$Se$_3$ interface at the end of the NbSe$_2$ flake. \textbf{c-d,} Atomic planes of NbSe$_2$ (c) and Bi$_2$Se$_3$ (d). 
%     }
%     \label{ext.TEM}
% \end{figure}

% \subsection*{Theoretical modeling}

% A LabView script was developed to numerically compute $I_s(\varphi_1, B)$ and $I_c(B)$ based on Eq. (2). The script is available in the Supplementary Information.

% \section*{   }

\section*{Keywords}
Josephson junctions, superconductivity, topological insulators, current-phase relation, superconducting diode effect, Majorana bound states.

\section*{Competing interests}
The authors declare no competing interests.

% \section*{Acknowledgements} 
% The work is supported by MOE Tier 2 grant Award T2EP50123-0020 (sample fabrication). We acknowledge the Electron Microscopy Facility (EMF) at the National University of Singapore for providing access to the FIB and STEM equipment. KSN is grateful to the Ministry of Education, Singapore (Research Centre of Excellence award to the Institute for Functional Intelligent Materials, I-FIM, project No. EDUNC-33-18-279-V12) and to the Royal Society (UK, grant number RSRP\ R\ 190000) for support. Z.L. acknowledges the support from National Research Foundation, Singapore, under its Competitive Research Programme  (CRP) (NRF-CRP22-2019-0007 and NRF-CRP22-2019-0004), and also the support by the Ministry of Education, Singapore, under its Research Centre of Excellence award to the Institute for Functional Intelligent Materials (Project No. EDUNC-33-18-279-V12).

% \section*{Author contributions}
% A.K. and D.A.B. designed and supervised the project. 
% X.Z. and A.K. fabricated the devices.
% A.K. performed the transport measurements.
% R.A.H. provided theoretical support.
% T.T. and K.W. provided the hBN crystals.
% L.A.Y provided the $\text{Bi}_2\text{Se}_3$ crystals.
% Xin.Z. performed STEM measurements.
% A.K. and D.A.B. wrote the manuscript with contributions from all authors.
% All authors contributed to the discussions. 

\bibliography{Bibliography.bib}

\newpage
\setcounter{figure}{0}
\renewcommand{\thesection}{}
\renewcommand{\thesubsection}{S\arabic{subsection}}
\renewcommand{\theequation} {S\arabic{equation}}
\renewcommand{\thefigure} {S\arabic{figure}}
\renewcommand{\thetable} {S\arabic{table}}
\end{document}

\setcounter{figure}{0}
\renewcommand{\thesection}{}
\renewcommand{\thesubsection}{S\arabic{subsection}}
\renewcommand{\theequation} {S\arabic{equation}}
\renewcommand{\thefigure} {S\arabic{figure}}
\renewcommand{\thetable} {S\arabic{table}}

\title{Supplementary Materials for \\  Non-Reciprocal Current-Phase Relation and Superconducting Diode Effect in Topological-Insulator-Based  Josephson Junctions}

% \title{Supplementary Materials for \\  Non-Reciprocal Current-Phase Relation and Superconducting Diode Effect in \\ Planar Bi$_2$Se$_3$/NbSe$_2$  Josephson Junctions}

\author{A. Kudriashov$^{+}$}
% \affiliation{Department of Materials Science and Engineering, National University of Singapore, 117575 Singapore}
\affiliation{Institute for Functional Intelligent
Materials, National University of Singapore, Singapore, 117575, Singapore}

\author{X. Zhou$^{+}$}
% \affiliation{Department of Materials Science and Engineering, National University of Singapore, 117575 Singapore}
\affiliation{Institute for Functional Intelligent
Materials, National University of Singapore, Singapore, 117575, Singapore}

\author{R.A. Hovhannisyan}
\affiliation{Department of Physics, Stockholm University, AlbaNova University Center, SE-10691 Stockholm, Sweden}

\author{A. Frolov}
\affiliation{Chemistry Department, M.V. Lomonosov Moscow State University, Moscow, Russian Federation}

\author{L. Elesin}
\affiliation{Institute for Functional Intelligent
Materials, National University of Singapore, Singapore, 117575, Singapore}
\affiliation{Programmable Functional Materials Lab, Center for Neurophysics and Neuromorphic Technologies, Moscow 127495}

\author{Y. Wang}
\affiliation{Institute for Functional Intelligent
Materials, National University of Singapore, Singapore, 117575, Singapore}
\author{E.V. Zharkova}
\affiliation{Programmable Functional Materials Lab, Center for Neurophysics and Neuromorphic Technologies, Moscow 127495}

\author{T.~Taniguchi}
\affiliation{International Center for Materials Nanoarchitectonics, National Institute of Material Science, Tsukuba 305-0044, Japan}

\author{K.~Watanabe}
\affiliation{Research Center for Functional Materials, National Institute of Material Science, Tsukuba 305-0044, Japan}

\author{Z. Liu}
\affiliation{School of Materials Science and Engineering, Nanyang Technological University, Singapore, 639798, Singapore}

\author{L.A. Yashina}
\affiliation{Chemistry Department, M.V. Lomonosov Moscow State University, Moscow}

\author{X. Zhou}
\affiliation{Institute for Functional Intelligent
Materials, National University of Singapore, Singapore, 117575, Singapore}

\author{K.S. Novoselov}
\affiliation{Institute for Functional Intelligent Materials, National University of Singapore, Singapore, 117575, Singapore}

\author{D.A. Bandurin*}
\affiliation{Department of Materials Science and Engineering, National University of Singapore, 117575 Singapore}

\maketitle

\tableofcontents

\subsection{Device fabrication}
In this section we describe in detail the fabrication of the device shown in the main text. 
Other devices were fabricated using similar technological processes.
\begin{figure*}[ht!]
    \begin{center}
    \includegraphics[width =0.8\columnwidth]{Stack.JPG}
    \caption{
    \textbf{Fabrication of the stack for the main device.}
    \textbf{a,} An optical micrograph of the hBN flake.
    \textbf{b,} An optical micrograph of the $\text{NbSe}_2$ flake.
    \textbf{c,} An optical micrograph of the $\text{Bi}_2\text{Se}_3$ flake.
    \textbf{d,} An optical micrograph of the final stack after the removal of melted PC.
}\label{stack}
    \end{center}
\end{figure*}

\textbf{$\text{NbSe}_2$/$\text{Bi}_2\text{Se}_3$/hBN heterostructure}

1) hBN was exfoliated at ambient conditions, using Nitto blue tape and $\text{Si/SiO}_2$ substrate.
Flake with appropriate size and thickness was determined using an optical microscope, as shown in Fig.\ref{stack}a.

2) $\text{NbSe}_2$ was exfoliated inside the Ar-field glovebox using Nitto blue tape and $\text{Si/SiO}_2$ substrate, heated up to 60-70°C. 
We were looking for flakes with a crack in it, as shown in Fig.\ref{stack}b.

3) $\text{Si/SiO}_2$ substrates for $\text{Bi}_2\text{Se}_3$ were cleaned in the oxygen plasma for 15 min (50 sccm of $\text{O}_2$, 250W, p=1.4e-2 mbar). 
$\text{Bi}_2\text{Se}_3$ flakes were exfoliated on the Nitto blue tape inside the glovebox and covered by another blue tape.
Then, these flakes on the covered tape were taken out of the glove box and the tape was placed in contact with freshly cleaned and hot substrate at ambient conditions.
Then covered by the blue tape substrate with flakes on it was put inside the glovebox and delaminated in the inert atmosphere. 
This process allows us to get thin $\text{Bi}_2\text{Se}_3$ flakes with freshly cleaved inside the glovebox surface.
It should be noted, that flakes were in contact with air for a few seconds before the tape was put in contact with a freshly cleaned substrate.
It may lead to slight oxidation of the bottom surface of the flakes.
After this, relatively thin flakes were identified under the optical microscope. We were looking for two thin flakes with different width close to each other to use them for fabrication of two Josephson junctions in the SQUID geometry.
The appropriate flakes are shown in Fig.\ref{stack}c.

4) hBN was picked up using PC/PDMS stamp at 90°C.
Then hBN/PC/PDMS stack was used to pick up $\text{NbSe}_2$ at 110°C.
In order to not deform the crack a lot, the front was perpendicular to the crack.
Then $\text{NbSe}_2$/BN/PC/PDMS was placed on top of the $\text{Bi}_2\text{Se}_3$ flakes so that the crack was perpendicular to the flake's edges and covered two flakes at the same time.
PC was melted at 180°C and washed away with DCM, followed by IPA.
The final stack is shown in Fig.\ref{stack}d.

\textbf{SQUID}

To fabricate the final device we used a combination of standard techniques, involving electron-beam lithography (EBL), reactive ion etching (RIE), and e-beam evaporation of the metals. The workflow of the device fabrication is shown in Fig.\ref{workflow}.

\begin{figure*}[ht!]
    \begin{center}
    \includegraphics[width =\columnwidth]{workflow.jpg}
    \caption{
    \textbf{Fabrication of the main device.}
    \textbf{a,} EBL (top panel) and deposition (lower panel) of the local flax line.
    \textbf{b,} EBL (top panel) and deposition (lower panel) of the contacts.
    \textbf{c,} EBL (top panel) and deposition (lower panel) of the top gates.
    \textbf{d,} EBL (top panel) and etching (lower panel) of the SQUID geimetry.
}\label{workflow}
    \end{center}
\end{figure*}

1) PMMA 495K A5 resist was spin-coated to the substrate and baked at 180°C for 2 min on a hotplate.
Then another layer of PMMA 950K A5 was spin-coated and baked at 180°C for another 2 min on a hotplate, forming a bilayer for creating under-cut.

2) EBL was performed to make markers for the precise positioning of the design with respect to the stack.

3) Deposition mask for the flux-line was created using EBL.

4) Mild 15 sec plasma (50 sccm of $\text{O}_2$, 50W, p=1.4e-2 mbar) was used to remove the resist residue. 
5 nm of Ti followed by 95 nm of Al were deposited using an e-beam evaporator with rates of 0.2 \r{A}/s and 0.1 \r{A}/s, respectively. 
After the lift-off in Acetone, the flux-line was fabricated.

6) Another PMMA bilayer was spin-coated similarly and EBL was used to create a mask for contacts.

7) hBN and $\text{NbSe}_2$ were etched for 30 sec using $\text{CHF}_3/\text{O}_2$ RIE with flows 20 sccm and 10 sccm, respectively, and RF power of 200 W. Temperature of the stage was kept at 10°C.

8) 5 nm of Ti followed by 95 nm of Au were deposited using an e-beam evaporator with rates of 0.2 \r{A}/s and 0.3 \r{A}/s, respectively. 
After the lift-off in Acetone, the edge contacts to $\text{NbSe}_2$ were fabricated. 

9) Another PMMA bilayer was spin-coated similarly and EBL was used to create a mask for the top-gates.

10)  Mild 15 sec plasma (50 sccm of $\text{O}_2$, 50W, p=1.4e-2 mbar) was used to remove the resist residue. 
5 nm of Ti followed by 95 nm of Au were deposited using an e-beam evaporator with rates of 0.2 \r{A}/s and 0.3 \r{A}/s, respectively. 
After the lift-off in Acetone, the top gates were fabricated.

11) Another PMMA bilayer was spin-coated similarly and EBL was used to create an etching mask for the SQUID patterning.

12) hBN and $\text{NbSe}_2$ were etched for 30 sec using $\text{CHF}_3/\text{O}_2$ RIE with flows 20 sccm and 10 sccm, respectively, and RF power of 200 W. Temperature of the stage was kept at 10°C.

13) After the resist removal in acetone, the device fabrication is finished.

\newpage
\textbf{From SQUID to two individual JJs}
\begin{figure*}[ht!]
    \begin{center}
    \includegraphics[width =0.7\columnwidth]{cut.jpg}
    \caption{
    \textbf{From SQUID3 to JJ1 and JJ2.}
    \textbf{a,} Optical photograph of the SQUID3 device discussed in the main text. 
    \textbf{b,} Two individual Josephson junctions JJ1 and JJ2 obtained after etching the SQUID into two not-connected devices.
}\label{cut}
    \end{center}
\end{figure*}

After the transport measurements of the SQUID3 (device from the main text), we cut it into two individual Josephson junctions in order to characterize them individually.

1) PMMA bilayer was spin-coated similarly and EBL was used to create an etching mask for the cuts.

2) The same etching recipe, as described above, was used to cut SQUID into two individual junctions, as shown in Fig.\ref{cut}.

\subsection{Further examples of the planal vdW Josephson junctions}

The first device which was fabricated using the process described in this work is shown in Fig.~\ref{devices}a.
It is a single Josephson junction, and its interference pattern is presented in Fig.~\ref{devices}b.
It displays a Fraunhofer-like dependence, similar to the devices JJ1 and JJ2 shown in the main text.
One notable difference is that JJ0 exhibits step-like features in its $I_c(B)$ dependence, along with hysteresis when the direction of the magnetic field sweep is reversed. We attribute this behavior to the presence of Abrikosov vortices, which significantly complicates the CPR measurements.
To solve this problem, we have observed that the etching of the $\text{NbSe}_2$ helps to prevent Abricosov vortices from influencing the measurements. 
It can be attributed to the creation of a potential barrier for vortex penetration or by the reduction of the total area of the superconductor, which increases the critical field at which vortices penetrate inside the superconductor\cite{stan2004}.

After successful demonstration of the proximity effect, we have fabricated SQUID, which is shown in Fig.~\ref{devices}c.
The oscillations of $I_c$ as a function of magnetic field $B$ for this device are shown in Fig.~\ref{devices}d. 
It demonstrates similar behavior as the device shown in the main text, namely, SQUID oscillations of $I_\text{c}$ demonstrate pronounced second harmonics and highly non-sinusoidal oscillations.
As this device was not equipped with a local magnetic flux line, an accurate measurement of the CPR was not feasible.

After that, we have fabricated two more SQUIDs with local flux line: SQUID2 and SQUID3. 
SQUID3 is the device from the main text, while SQUID2 was used to perform detailed STEM analysis, demonstrated in Extended figure in the main text and below. 
\begin{figure*}[ht!]
    \begin{center}
    \includegraphics[width =0.8\columnwidth]{devices.png}
    \caption{\textbf{Other devices.} 
    \textbf{a,} Optical micrograph of the sample named JJ0. Green and yellow dashed lines highlight $\text{NbSe}_2$ and $\text{Bi}_2\text{Se}_3$ flakes, respectively.
    \textbf{b,} Critical current $I_{\text{c}}$ and retrapping current $I_{\text{r}}$ of device JJ0 plotted versus magnetic field $B$.
    \textbf{c,} Optical micrograph of the sample named SQUID1. Green and yellow dashed lines highlight $\text{NbSe}_2$ and $\text{Bi}_2\text{Se}_3$ flakes, respectively.
    \textbf{d,} Critical current $I_{\text{c}}$ of device SQUID1 (blue line) plotted versus magnetic field $B$. Red lines show the envelope functions. The inserts display zoomed-in views of the SQUID oscillations near the minima of the interference pattern.
}\label{devices}
    \end{center}
\end{figure*}

\newpage

\subsection{STEM and EDS}
The STEM image of two JJs are shown in Fig.~\ref{TEM}a. 
As one can see, the distance between two $\text{NbSe}_2$ flakes (a crack size) is about 150 nm, and the height of hBN is about 40 nm. 
There is some 5 nm thick contrast region where $\text{Bi}_2\text{Se}_3$ is supposed to be, however, we could not get an atomic resolution.
Most likely, it happened because the cut was done in a slightly wrong place: right on the edge of $\text{Bi}_2\text{Se}_3$. 
The thickness of $\text{NbSe}_2$ flake is about 12 nm, as shown in Fig.\ref{TEM}b.

\begin{figure*}[ht!]
    \begin{center}
    \includegraphics[width =\columnwidth]{TEM_sup.jpg}
    \caption{
    \textbf{STEM and EDS analysis}
    \textbf{a,} STEM of the junction region of JJ2. 
    \textbf{b,} HAADF-STEM of the $\text{NbSe}_2$ region.
    \textbf{c,} STEM image of the junction region of SQUID2 near the edge of the $\text{NbSe}_2$ crack. 
    \textbf{d-h,} Energy dispersive spectroscopy 
(EDS) mapping of the same region showing distributions of Nitrogen, Silicon, Selenium, Bismuth, and Niobium, respectively.
}\label{TEM}
    \end{center}
\end{figure*}

The detailed STEM analysis of the sample SQUID2 is shown in the Fig.~1 of the main text. Elemental distribution in the junction near the edge of the $\text{NbSe}_2$ crack is shown in Fig.~\ref{TEM}c-h.

% \begin{figure*}[ht!]
%   \centering\includegraphics[width=0.8\linewidth]{Extended TEM.JPG}
%     \caption{\textbf{Scanning transmission electron microscopy analysis of the vdW JJ. a,} Typical STEM image of the device cross section. \textbf{b, } Zoomed-in STEM image of the NbSe$_2$/Bi$_2$Se$_3$ interface at the end of the NbSe$_2$ flake. \textbf{c-d,} Atomic planes of NbSe$_2$ (c) and Bi$_2$Se$_3$ (d). 
%     }
%     \label{ext.TEMSI}
% \end{figure*}

\newpage

\subsection{Estimation of $\xi_\text{n}$}
The coherence length ($\xi_\text{n}$) in $\text{Bi}_2\text{Se}_3$ can be estimated in two limiting cases: the clean limit (when the mean free path $l$ is much larger than the coherence length) and the dirty limit (when $l$ is much shorter than the coherence length).
In the clean limit, the coherence length is given by:
\begin{equation}
\xi_\text{n} = \frac{\hbar v_\text{F}}{2\pi k_\text{B} T},
\end{equation}
where $v_\text{F}$ is the Fermi velocity, $k_\text{B}$ is the Boltzmann constant, and $T$ is the temperature. At $T = 10$~mK, this yields $\xi_\text{n} \simeq 12~\mu\text{m}$.

In the dirty limit, the coherence length becomes:
\begin{equation}
\xi_\text{n} = \sqrt{\frac{\hbar v_\text{F} l}{6\pi k_\text{B} T}},
\end{equation}
where $l$ is the mean free path. For typical values of $l \sim 10\text{--}100$~nm in our $\text{Bi}_2\text{Se}_3$ samples, we estimate $\xi_\text{n} \simeq 0.45\text{--}1.4~\mu\text{m}$. Given that our junction length $L$ is much smaller than both estimated values of $\xi_\text{n}$, the device operates in the short junction regime ($L \ll \xi_\text{n}$).

\subsection{Interference pattern for varying $W/\lambda_J$}
To test whether our Josephson junction is in the wide ($\lambda_J<W$, $\lambda_J$ - Josephson penetration length) or narrow regime ($\lambda_J>W$), we compare experimentally measured $I_\mathrm{c}(B)$ with that calculated for junctions of varying width $W$. To this end, we use sine-Gordon equation~\cite{barone}:  
\begin{equation}\label{SinGordon}
    \frac{\partial^2\varphi}{\partial \bar{y}^2}  = \sin(\varphi) - j_b, 
\end{equation}  
with normalized $\bar{y} = y / \lambda_J$. Here \( j_b \) represents the bias current, while \( \varphi \) denotes the phase difference at the coordinate \( y \). Assuming standard boundary conditions  
\begin{equation}
\dfrac{d\varphi}{d\bar{y}}\bigg|_{\bar{y}=\pm W/2} = \dfrac{2\pi}{W}\frac{\Phi_{1}}{\Phi_0},
\end{equation}
we obtain $I_\mathrm{c}(B)$ for different ratios of $W/\lambda_J$ presented in Fig.~\ref{WideJJ}. Here $\Phi_{1}$ is the magnetic flux through JJ1 and $\Phi_0$ is the flux quantum. The two extreme cases $W/\lambda_J=0$ and $W/\lambda_J=5$ can be distinguished by the notably different functional forms of $I_\mathrm{c}(B)$ in the vicinity of zero $B$. In particular, at small $B<0.75~$mT and when $W=0$, the $I_\mathrm{c}(B)$ dependence reproduces familiar Fraunhofer pattern $I_c(\Phi) = I_c(0) \left| \frac{\sin(\pi \Phi / \Phi_0)}{\pi \Phi / \Phi_0} \right|$. On the contrary, the zero-order Fraunhofer peak is significantly wider for the of $W/\lambda_J=5$ so that even the first Fraunhofer minimum is obscured. When compared to the experiment, the width of the measured $I_\mathrm{c}(B)$ peak around $B=0$ follows closely the $W=0$ model indicating that our JJ1 being in the narrow regime. As explained in the main text, the overall deviation of the $I_\mathrm{c}(B)$ from the standard Fraunhofer pattern originates from the non-uniform supercurrent distribution. 

\begin{figure}[h!]
    \begin{center}
    \includegraphics[width = 0.5\columnwidth]{WideJJ.pdf}
    \caption{
    \textbf{Interference pattern for varying $W/\lambda_J$}. Symbols: Experimental $I_\mathrm{c}(B)$ for JJ1. Solid lines: Calculated $I_\mathrm{c}(B)$ dependecies for given $W/\lambda_J$.
}
    \label{WideJJ}
    \end{center}
\end{figure}

% As one can see, the experimental data for JJ1 is better fitted by the model for a narrow junction ($W=0$), which is a standard Fraunhofer pattern $I_c(\Phi) = I_c(0) \left| \frac{\sin(\pi \Phi / \Phi_0)}{\pi \Phi / \Phi_0} \right|$. 
% This proves that our JJ1 is not in a wide regime, therefore it does not explain the node lifting. 

\newpage
\subsection{Current source offset compensation}
% For aquarate measurements of the superconducting diode effect it is crucial to know the exact value of the applied current. 
% However, a DC offset is inevitably present in all current sources due to technical limitations.
% To measure and compensate for a current offset of the voltage-controlled current source CS580, the following procedure was implemented.
% A 1 $\text{M}\Omega$ resistor was connected to the output of the voltage-controlled current source CS580 with input and output disabled. The voltage drop on the resistor was amplified with the SR560 voltage amplifier and the amplified voltage was measured using NI-3232, as shown in the inset of Fig.~\ref{offset}. 
% Since the voltage amplifiers also have DC offset, we subtracted it from the further readings (because in this configuration no current can flow, therefore the real voltage drop on the resistor should be zero).

For accurate measurements of the superconducting diode effect, it is crucial to know the exact value of the applied current.
However, a DC offset is inevitably present in all current sources due to technical limitations.
To measure and compensate for the current offset of the voltage-controlled current source CS580, the following procedure was implemented.
A 1 $\text{M}\Omega$ resistor was connected to the output of the voltage-controlled current source CS580, with both input and output disabled. The voltage drop across the resistor was amplified using the SR560 voltage amplifier, and the amplified voltage was measured using the NI-3232, as shown in the inset of Fig.~\ref{offset}.
Since the voltage amplifiers also exhibit a DC offset, this offset was subtracted from subsequent readings (because in this configuration, no current can flow, and therefore, the actual voltage drop across the resistor should be zero).
  
Fig.~\ref{offset} shows the measured current flowing through 1 $\text{M}\Omega$ resistor. 
As one can see, when the output of CS580 is disabled, no current is flowing. 
However, when the output was enabled at approximately 40 seconds after the beginning of the experiment, a finite current of about 2 nA started to flow. 
Current increases to 32 nA when the input (with QDevil qDAC-II connected to it) is enabled at approximately 65 sec.
\begin{figure}[h!]
    \begin{center}
    \includegraphics[width = 0.4\columnwidth]{offset.png}
    \caption{
    \textbf{Offset compensation.}
    Measured value of output current as a function of time. Inset shows the measurement circuit.  
}
    \label{offset}
    \end{center}
\end{figure}
To compensate for this finite current, -32 nA DC current was added to the output using the offset compensation function of the CS580.

\bibliography{Bibliography.bib}

\newpage
\setcounter{figure}{0}
\renewcommand{\thesection}{}
\renewcommand{\thesubsection}{S\arabic{subsection}}
\renewcommand{\theequation} {S\arabic{equation}}
\renewcommand{\thefigure} {S\arabic{figure}}
\renewcommand{\thetable} {S\arabic{table}}